\documentclass[showpacs,aps,prb,reprint,superscriptaddress]{revtex4-2}
\usepackage{bm}
\usepackage{graphicx}
\usepackage{amsmath}
\usepackage{amsfonts}
\usepackage{amssymb}
\usepackage{epstopdf}
\usepackage{hyperref}
\usepackage{float}
\hypersetup{
    colorlinks,%
    citecolor=blue,%
    linkcolor=blue,%
    urlcolor=blue
}
\usepackage{tikz}\usetikzlibrary{petri}

\newcommand{\kmpar}{\lambda_{\mathrm{SO}}}

\begin{document}

    \newcommand{\be}   {\begin{equation}}
    \newcommand{\ee}   {\end{equation}}
    \newcommand{\ba}   {\begin{eqnarray}}
    \newcommand{\ea}   {\end{eqnarray}}
    \newcommand{\ve}  {\varepsilon}
    
    \title{ Enhanced Majorana bound states in magnetic chains on superconducting topological insulator edges }
        
    \author{Raphael L.~R.~C. Teixeira}
    \affiliation{Instituto de F\'{\i}sica, Universidade de S\~{a}o Paulo,
        C.P.\ 66318, 05315--970 S\~{a}o Paulo, SP, Brazil}
    \author{Dushko Kuzmanovski}
\affiliation{Nordita, KTH Royal Institute of Technology, and Stockholm University SE-106 91 Stockholm, Sweden}
    \affiliation{Department of Physics and Astronomy, Uppsala University, Box 516, SE-751 20 Uppsala, Sweden}
    \author{Annica M.~Black-Schaffer}
    \affiliation{Department of Physics and Astronomy, Uppsala University, Box 516, SE-751 20 Uppsala, Sweden}
    \author{Luis G.~G.~V. Dias da Silva}
    \affiliation{Instituto de F\'{\i}sica, Universidade de S\~{a}o Paulo,
        C.P.\ 66318, 05315--970 S\~{a}o Paulo, SP, Brazil}
    
    \date{ \today }
    
    \begin{abstract}
     The most promising mechanisms for the formation of Majorana bound states (MBSs) in condensed matter systems involve one-dimensional systems such as semiconductor nanowires, magnetic chains, and quantum spin Hall insulator (QSHI) edges proximitized to superconducting materials. The choice between each of these options involves trade-offs between several factors such as  reproducibility of results, system tunability, and robustness of the resulting MBS. In this paper, we propose that a combination of two of these systems, namely, a magnetic chain deposited on a QSHI edge in contact with a superconducting surface, offers a better choice of tunability and MBS robustness compared to magnetic chain deposited on bulk. We study how the QSHI edge interacts with the magnetic chain, and see how the topological phase is affected by edge proximity. We show that MBSs near the edge can be realized with lower chemical potential and Zeeman field than the ones inside the bulk, independently of the chain's magnetic order (ferromagnetic or spiral order). Different magnetic orderings in the chain modify the overall phase diagram, even suppressing the boundless topological phase found in the bulk for chains located at the QSHI edge. Moreover, we quantify the ``quality" of MBSs by calculating the Majorana polarization (MP) for different configurations. For chains located at the edge, the MP is close to its maximum value already for short chains. For chains located away from the edge, longer chains are needed to attain the same quality as chains located at the edge. The MP also oscillates in phase with the in-gap states, which is relatively unexpected as peaks in the energy spectrum correspond to stronger overlap of MBSs.
    \end{abstract} 
    \maketitle

    \section{Introduction}
    \label{sec:Intro}
  
Recent theoretical proposals and experimental efforts to implement topological 
qubits based on non-Abelian anyons~\cite{Kitaev:P.U:2001,Nayak:Rev.Mod.Phys.:1083--1159:2008,Fornieri:Nature:2019} have attracted a great deal of attention to the field of topological superconductivity. One of the simplest realizations of anyons with non-Abelian statistics are Majorana bound states (MBSs), which naturally emerge as edge states in one-dimensional (1D) $p$-wave topological superconductors~\cite{Alicea:Reports:2012,Aguado::40:523-593--:2017}. Several experimental efforts have been devoted to finding systems that can host MBSs. Examples include magnetic chains on superconducting surfaces~\cite{Nadj-Perge:Phys.Rev.B:020407:2013,Nadj-Perge:Science:602--607:2014}, topological insulators proximitized with superconductors~\cite{Fu:Phys.Rev.Lett.:96407:2008,Inglot:JournalofAppliedPhysics:123709:2011,Gonzalez:Phys.Rev.B:115327:2012,Kuzmanovski:PhysRevB.94.180505,He:Phys.Rev.B:075126:2013}, and semiconductor nanowires with strong spin-orbit coupling close to a superconductor~\cite{Oreg:Phys.Rev.Lett.:177002:2010,Lutchyn:Phys.Rev.Lett.:77001:2010}

When it comes to the first example system above, Ref.~\cite{Nadj-Perge:Phys.Rev.B:020407:2013} 
put forward the idea that magnetic adatoms, forming a 1D chain and deposited on top of a superconducting surface, could give rise to MBSs. Here, we refer to this arrangement as ``MAG+SC.'' The interplay of the coupling between the localized magnetic moments and superconductivity leads to a regime where the chain behaves as a 1D topological superconductor with MBSs appearing at its edges. Several experiments have reported the presence of localized states at the end of magnetic chains, consistent with MBSs' signature~\cite{Nadj-Perge:Science:602--607:2014,Ruby:Phys.Rev.Lett.:197204:2015,Pawlak:NpjQuantumInformation:2:16035:2016,Jeon:Science:772:2017,Kim:ScienceAdvances::2018}. 
Due to its simplicity, the idea behind ``MAG+SC'' has been used in different contexts to understand the behavior of MBSs. However, the magnetic impurities themselves are not the only important element for the formation of MBSs, and different substrates can change the topological phase.

One of the first proposals for topological superconductivity was in fact to use topological insulator as a substrate. Fu and Kane~\cite{Fu:Phys.Rev.Lett.:96407:2008} considered the case of BCS superconductors on the surface of a 3D topological insulator, where a magnetic domain produces chiral, propagating Majorana modes around the domain. Reducing the spatial dimension to a 2D topological insulator, also known as a quantum spin Hall insulator (QSHI), generates automatically bound Majorana states, i.e.,~MBSs, at junctions between magnetism and superconductivity. We hereafter dub this proposal ``QSHI+SC+FM,'' which consists of proximitizing the edge of a QSHI with an $s$-wave superconductor (such that the edge modes' dispersion is gapped), and forming a 1D junction using a ferromagnetic insulator~\cite{Fu_Phys.Rev.B_161408_2009}. This arrangement gives rise to MBSs at the end-points of the junction, where the three materials meet~\cite{Alicea:Reports:2012}. The possibility of using a QSHI as a substrate opens many routes to study MBSs.

Early theoretical proposals for the realization of the QSHI phase involves electrons in a honeycomb lattice and strong spin-orbit interaction~\cite{Kane:Phys.Rev.Lett:226801:2005}. Experimental studies have also shown that some honeycomb lattice materials can display QSHI edge states. Recent examples include graphene decorated with Bi$_2$Te$_3$~\cite{Hatsuda:AAAS:2018} and monolayer WTe$_2$ systems~\cite{Wu:Science:76:2018}. Other studies have shown that graphene-like materials, such as silicene and stanene, even host induced superconductivity when doped~\cite{Chen_2013,Spencer:Springer:2016, Kuzmanovski:PhysRevB.94.180505,Teixeira:PhysRevB.99.035127:2019}. Together, these properties make two-dimensional (2D) topological insulators with honeycomb geometry a promising candidate for realizing MBSs in the QSHI+SC+FM scheme. 

In a previous work~\cite{Teixeira:PhysRevB.99.035127:2019}, we established that 
a combination of the QSHI+SC+FM and MAG+SC approaches 
can realize MBSs at the ends of a magnetic chain placed in the \emph{bulk}
of a QSHI with induced superconductivity. This arrangement allows for phase diagrams with ``boundless'' topological phases, where the topological phase is independent on certain parameters, but, notably, where the form of which depends crucially on the magnetic ordering in the chain.  

Here, we take one decisive step further and utilize the most defining property of QSHI, the edge state. More specifically,
we explore  the non-trivial role the QSHI \emph{edge states} play in the stability and robustness of the MBS located at the ends of a magnetic chain. 
Concretely, we study a chain of magnetic adatoms deposited on a QSHI described by the Kane-Mele model on a zigzag strip of honeycomb lattice with induced superconductivity. This way, it is possible to probe the edge-Majorana interaction just by moving the chain either toward or away the edge and changing the magnetic order.

We show that the coupling of the magnetic chain with QSHI edge states has several important consequences for the formation and stabilization of MBSs. First, we find that the shape of the MBS-hosting topological region in the doping vs.~magnetic-impurity-strength phase diagram depends not only on the magnetic ordering along the chain but changes rather strongly depending on the chain position relative to the QSHI edge. More interestingly, we find that this effect is restricted to the edge region, extending only over distances of about four times the lattice constant, i.e., two full hexagons. 
The strength of the superconducting order parameter does not change any qualitative properties of the topological phase, suggesting that it is the interaction between QSHI edge states and the  magnetic chain that is crucial for the distinct features we observe. This fact is the reason why the MBSs behave very differently between bulk and edge locations. This is further supported by the fact that when the substrate is in a topological insulator phase, the MBSs behavior depends on its position. However, when the substrate is in a normal phase, we do not observe any drastic differences between MBSs in different positions.

In addition, we are able to evaluate the ``quality'' of the MBSs by calculating the ``Majorana polarization'' (MP) introduced 
in Refs.~\cite{Sticlet:PhysRevLett.108.096802:2012,Sedlmayr:PhysRevB.92.115115:2015}. The MP gives a quantitative measure of the superposition of electron and hole states forming  MBSs: in other words, the MP is maximum for a ``pure,'' particle-hole-symmetric MBS formed by equal contributions of electron and hole states. 
Although finite-chain-length effects can lead to a decrease in the MP (due to the coupling between MBSs located at the different extremes of the chain), our results show that the MP increases significantly when the chain is located at the QSHI edge, thus strongly reducing any finite-size effects. Another way to put this is that, for a given chain length, we obtain sharper, more localized, MBSs by placing the chain at the QSHI edge rather than in the QSHI bulk. By contrast, non topological Andreev bound states (ABSs) do not show such dependence due to their non localized nature. This different behavior adds the intriguing prospect of using the distance of the chain to a QSHI edge to differentiate MBSs and ABSs states in experiments.

 The remainder of the text is organized as follows: In Sec.~\ref{sec:modelmethods} we introduce the model Hamiltonian used to describe the system and discuss the methods used throughout this paper. The phase diagram for the system is studied in Sec.~\ref{sec:TopPhase}, where we classify the topological phase and discuss its dependence with the spiral magnetic angle, doping and distance to the edge. 
 The interplay of MBSs and QSHI edge states is further explored in Sec.~\ref{sec:MBS}. There, we focus on the dependence of the low-energy spectrum (and the MBS) with the position of the chain relative to the QSHI edge. We also show the dependence of the MP with the size, spiral magnetic angle and local magnetic moments of the chain. Finally, we present our concluding remarks in Sec.~\ref{sec:Conclusions}.

  \section{Model and Methods}
  \label{sec:modelmethods}
 
\begin{figure}[t]
    \begin{center}
        \includegraphics[width=1\columnwidth]{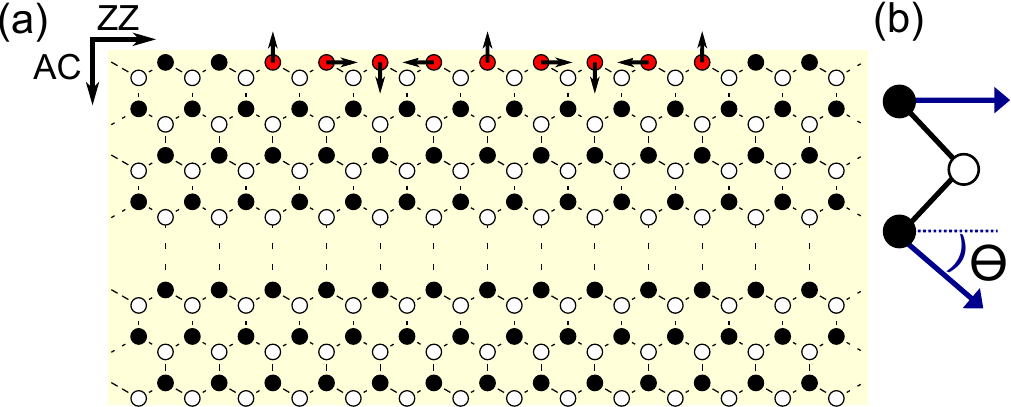}
        \caption{(a) Honeycomb lattice on top of a superconducting surface (yellow) with white and black dots marking the two sublattices. Dotted lines in the middle indicate the bulk. Magnetic impurities are represented as sites in red with spiral magnetic order, $\theta=\pi/2$.  Zigzag edge (ZZ) and armchair (AC) directions are also marked. (b) Angle $\theta$ between the neighboring magnetic moments defines the magnetic spiral  with in-plane magnetic moments.} 
        \label{fig:model}
    \end{center}
\end{figure}

A sketch of the system is shown in Fig.~\ref{fig:model}(a). A horizontal strip  of a honeycomb lattice material with a zigzag top edge is deposited on a superconducting surface (yellow background). The sites of the two sublattices are represented by black and white dots, while the magnetic impurity chain of adatoms placed at the edge are represented in red, with arrows representing each magnetic moment in the $xy$-plane. Each impurity has its magnetic moment rotated by an angle $\theta$ from the preceding one, as shown in Fig.~\ref{fig:model}(b). In this work, we consider the cases $\theta=0$ (indicating ferromagnetic order), and $\theta=\pi/2$ (spiral magnetic order), as illustrated in Fig.~\ref{fig:model}(a).

We assume that the honeycomb lattice material can be described by the Kane-Mele model~\cite{Kane:Phys.Rev.Lett:226801:2005},  allowing for the appearance of QSHI behavior and topologically-protected helical edge states. As such, the system is described by the complete  Hamiltonian $\mathcal{H}=\mathcal{H}_{\rm KM} + \mathcal{H}_{\rm SC} + \mathcal{H}_{\rm imp}$, where
\begin{subequations}
    \begin{eqnarray}
    & \mathcal{H}_{\rm KM} = t \, \sum_{\langle i,j \rangle, \sigma}c^{\dagger}_{i,\sigma} \, c_{j,\sigma} \nonumber \\
    & + i \, \frac{\kmpar}{3 \sqrt{3}} \, \sum_{\langle\langle i,j \rangle\rangle, \sigma} \nu_{ij} \, c^{\dagger}_{i,\sigma} \, (s_{z})_{\sigma\sigma'} \, c_{j,\sigma'} \nonumber \\
    & - \mu \sum_{i} c^{\dagger}_{i, \sigma} c_{i,\sigma},\label{eq.KM}, \\
    & \mathcal{H}_{\rm SC} = \sum_{i} \left\lbrace \Delta_{i} \, c^{\dagger}_{i,\uparrow} \, c^{\dagger}_{i,\downarrow} +{\rm H.c.} \right\rbrace, \\
    & \mathcal{H}_{\rm imp}= \sum_{i\in\mathcal{I}, \sigma, \sigma'} V_z \, c^{\dagger}_{i,\sigma} \, (\hat{n_{i}}\cdot \vec{s})_{\sigma\sigma'} \, c_{i,\sigma'}\label{eq.Imp} \; .
    \end{eqnarray}
\end{subequations}

In the above, $\mathcal{H}_{\rm KM}$ is the Kane-Mele Hamiltonian, in which $t$ is the nearest--neighbor hopping in the honeycomb lattice, $\mu$ is the on-site chemical potential, and $\kmpar$ is the spin-orbit coupling strength between next-nearest-neighbors within each sublattice. The spin-orbit chirality is given by $\nu_{ij}= (\bm{d}_{i} \times \bm{d}_{j})_{z}=\pm 1$, with $\bm{d}_{i,j}$ the unitary vectors connecting sites $i$ and $j$, and $s_i$ is the vector of Pauli matrices in spin space. 

Proximity-induced superconductivity is modeled via a BCS-type Hamiltonian $\mathcal{H}_{\rm SC}$, where $\Delta_{i}$ is a spin-singlet $s$-wave superconducting order parameter. It is determined through the self-consistency condition:
\begin{equation}
\label{eq:SelfConsCond}
\Delta_{i} = U_{\rm sc} \, \left\langle c_{i,\uparrow} \, c_{i,\downarrow} \right\rangle.
\end{equation}
This emulates the leaking of Cooper pairs from the superconductor into the QSHI via an effective electron-electron interaction characterized by an on-site attraction with strength $-U_{\rm sc}$~\cite{BlackSchaffer:PhysRevB.78.024504, BlackSchaffer:PhysRevB.82.184522, BlackSchaffer:PhysRevB.83.220511}.
The condition Eq.~(\ref{eq:SelfConsCond}) is updated iteratively until two steps have an absolute difference of less than $10^{-3}$ and is calculated for a clean sample, i.e. no impurities. 

Finally, the magnetic impurities are added to the system through the Hamiltonian  $\mathcal{H}_{\rm imp}$, where the summation runs over the subset of sites $i \in \mathcal{I}$ holding adatoms. We describe the magnetic impurities as independent spins, with a Zeeman-type strength  $V_z$ favoring in-plane alignment at a direction $\hat{n_{i}}(\theta_i) =\left( \cos\left[\theta_i \right],\sin\left[\theta_i \right],0\right)$. Here, $\theta_i=\theta x_i/L$ is the alignment angle for a impurity at a lattice index position $x_i$  in the chain of length $L$ and  $\theta$ is the \emph{spiral angle}, which defines the type of magnetic order along the chain. For instance, $\theta\!=\!0$ and $\theta\!=\!\pi$ correspond, respectively, to  \emph{ferromagnetic} and \emph{anti-ferromagnetic} order in the chain, while intermediate values between those indicate spiral magnetic order.

Throughout this work, we use $\lambda_{\rm SO}=0.5t$ (which gives a normal-state full energy gap 2$\lambda_{\rm SO}=t$) and an effective superconducting attraction  $U_{\rm sc}=2t$. These parameters yield a superconducting order parameter $\Delta_b\sim 10^{-3}t$, in the bulk, and $\Delta_e\sim 3\times10^{-1}t$ at the edges, for small doping levels. Due to the self-consistent condition, the order parameter in the bulk increases with $\mu$, while the dependence at the edge with $\mu$ is much weaker, due to the approximately constant density of states of the edge states. As $\mu$ increases, the normal-state energy spectrum goes from that of a  topological insulator to that of a metal, at $\mu=\kmpar$. Close to the metallic phase, we observe a strong increase in $\Delta$ at the bulk, which continues with increasing $\mu$. This, however, does not affect the existence of a topological phase, as we find that  fixing a constant $\Delta$ for all sites does not change the phase diagram, although general properties of the system do change.
The superconducting order in the middle of the sample is compatible to the one found in the full bulk calculations~\cite{Teixeira:PhysRevB.99.035127:2019}, i.e.~not using a finite slab system, which ensures that the bulk properties are the same.

We obtain the spectrum of $\mathcal{H}$ by solving the Bogoliubov-de Gennes equations numerically. We use both periodic boundary conditions (PBCs) and open boundary conditions (OBCs) along the  zigzag edge direction, where the magnetic chain is located. For PBCs, the magnetic chain with $N_x$ sites spans the full length of the strip. We then choose $N_x$ such that we have a integer number of spiral rotations, with $k$ given by $N_x = 2\pi k/\theta$ where $\theta$ is the spiral angle. Alternatively, we simulate finite chains by using OBCs, with the chain fully embedded in the host. In this case, we use a supercell approach, such that there is a large buffer region isolating two copies of the magnetic chain. Although there is no constraint on $\theta$, we use the same values as for the PBC calculations to ease comparisons. We also impose that the first and last impurity have the same magnetic alignment. In all cases, the transverse size (armchair), remains fixed with $N_y=60$ sites in order to ensure stable bulk conditions in the interior of the QSHI.

 \section{Topological phase diagrams}
  \label{sec:TopPhase}
We start by establishing the topological phase diagrams as function of both doping and strength of the magnetic impurities.
The existence of topological phases and the phase diagram can be determined by the ``Majorana number''~\cite{Kitaev:P.U:2001,Teixeira:PhysRevB.99.035127:2019} for the system with PBCs. As mentioned previously, in this case the magnetic chain spans the full length of the strip of $N_x$ sites. We then use the definition of Majorana number in terms of Pfaffian (Pf)
   \begin{equation}
 \label{eq:M}
 \mathcal{M(H_A)}=\frac{\mbox{Sgn } \left[\mbox{Pf }\left(\mathcal{H_A}(N_1+N_2)\right)\right]}{\mbox{Sgn }\left[\mbox{Pf }\left(\mathcal{H_A}(N_1)\right)\right]\mbox{Sgn }\left[\mbox{Pf }\left(\mathcal{H_A}(N_2)\right)\right]},
 \end{equation}
   where $\mathcal{H_A}$ is the antisymmetric form of the Hamiltonian $\mathcal{H}$ \cite{Kitaev:P.U:2001}, 
   \begin{equation}
       \mathcal{H_A}=\frac{1}{2}\begin{pmatrix}
1 & 1\\
i & -i
\end{pmatrix} \mathcal{H}
\begin{pmatrix}
1 & -i\\
1 & i
\end{pmatrix}
\end{equation}
with $\mathcal{H}$ given by Eq.~\eqref{eq.KM}-\eqref{eq.Imp} in the basis $(c^\dagger_\uparrow,c^\dagger_\downarrow,c_\uparrow,c_\downarrow)$, Sgn is the sign function and $N_1$ and $N_2$ are the lengths of two different chains. Assuming $N_1\!=\!N_2$ we can simplify the expression to $\mathcal{M(H)}=\mbox{Sgn} \left[\mbox{Pf }\left(\mathcal{H}(2N_1)\right)\right]$ and the Majorana number can thus be determined by a calculation of the Pfaffian of the Hamiltonian for a chain of $2N_1$ sites.  Thus, for a given spiral angle $\theta$, we use only chains with an even number of spiral rotations $k$ given by $N_x = 2\pi k/\theta$ such that we only need to calculate the Pfaffian once (see Ref.~\onlinecite{Teixeira:PhysRevB.99.035127:2019} for details).

\begin{figure}[t]
    \begin{center}
        \includegraphics[width=1\columnwidth]{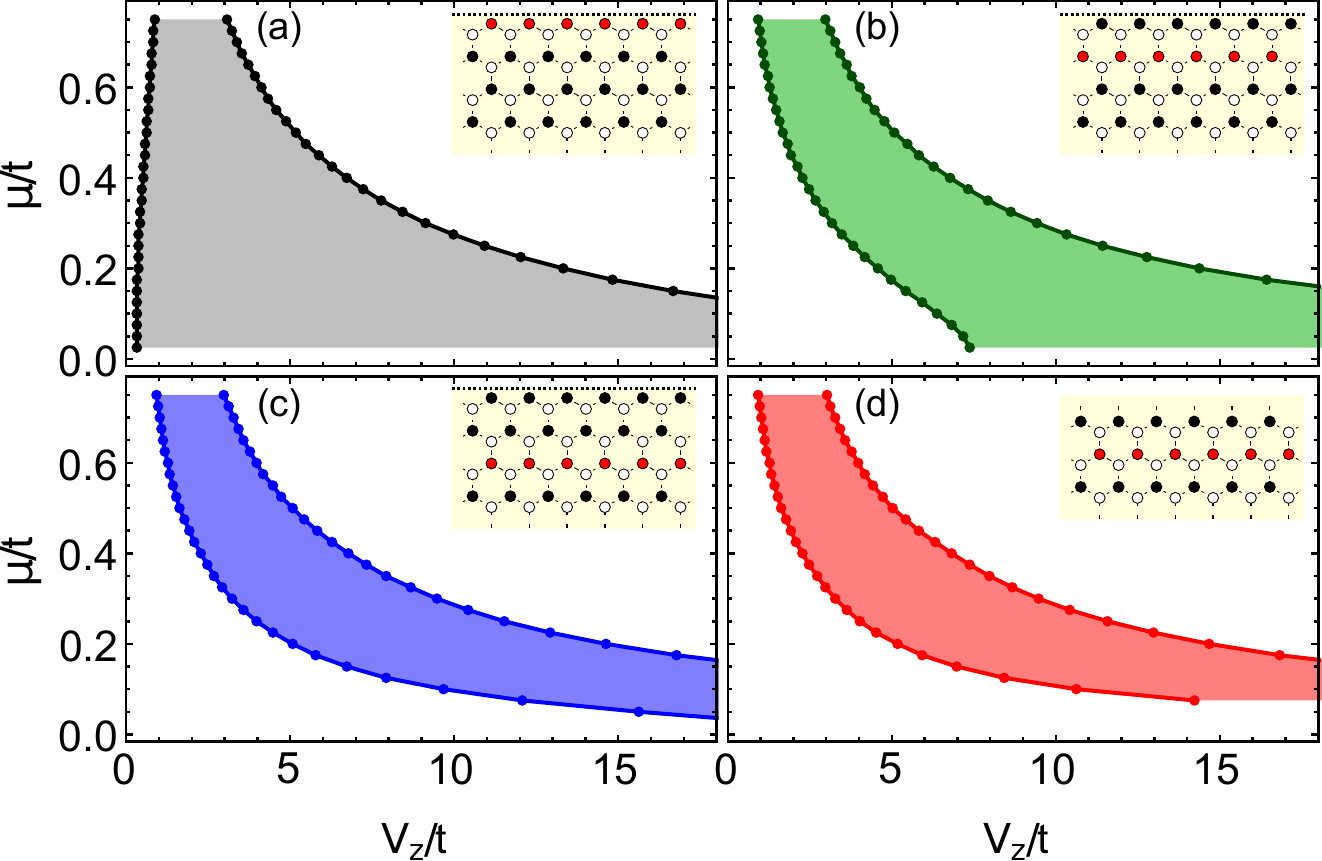}
        \caption{Phase diagram for a ferromagnetic chain ($\theta=0$). Topologically nontrivial phases ($\mathcal{M}=-1$) are shaded in (a) black for a chain at the edge, (b) green for a chain $3a/2$ away from the edge, (c) blue for a chain $3a$ away from the edge and (d) red for a chain  in the bulk.}
        \label{fig:phase0}
    \end{center}
\end{figure}

\begin{figure}[t]
    \begin{center}
        \includegraphics[width=1\columnwidth]{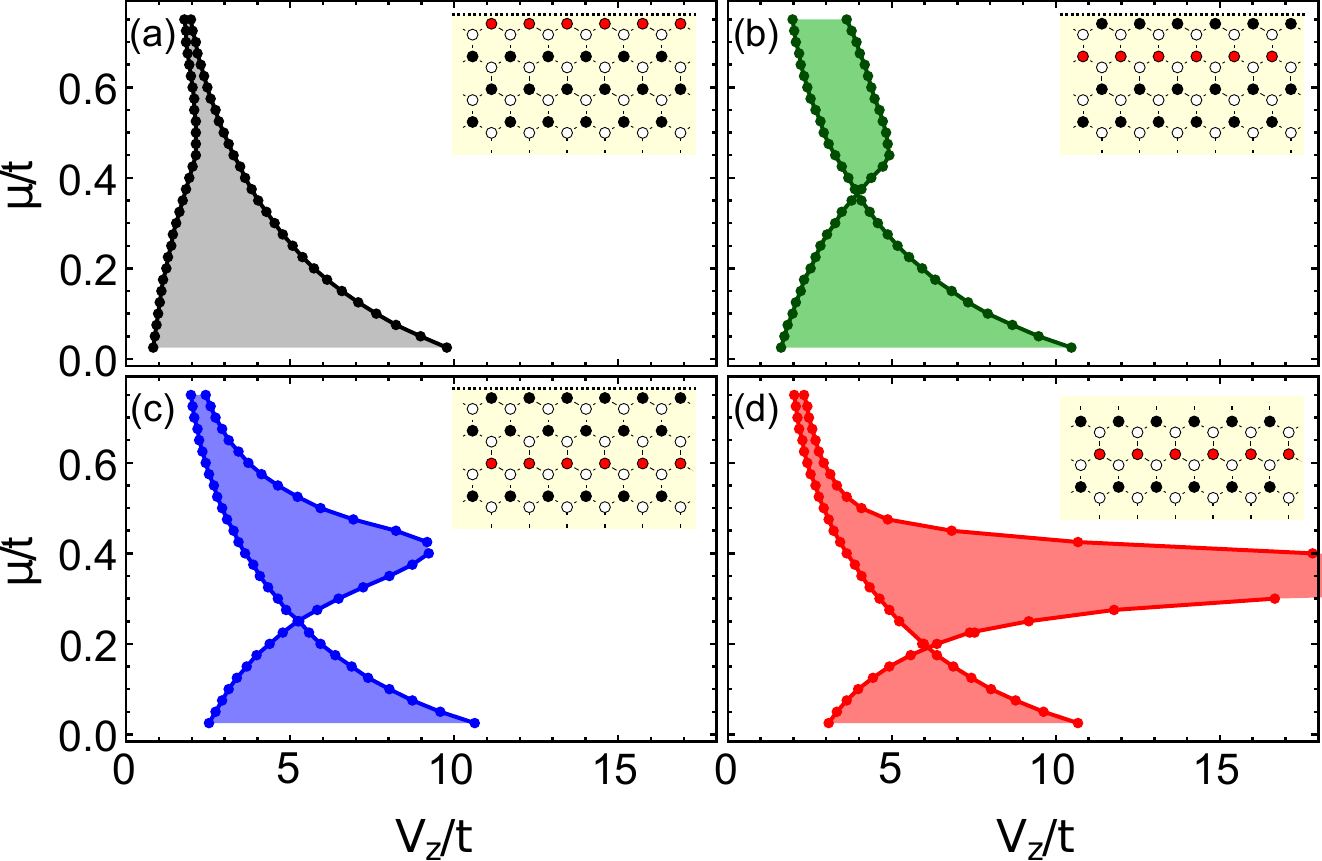}
        \caption{Phase diagram for a spiral chain ($\theta=\pi/2$). Topologically nontrivial phases ($\mathcal{M}=-1$) are shaded in (a) black for a chain at the edge, (b) green for a chain $3a/2$ away from the edge, (c) blue for a chain $3a$ away from the edge and (d) red for a chain  in the bulk.}
        \label{fig:phasePi2}
    \end{center}
\end{figure}

 The phase diagrams, Figs.~\ref{fig:phase0} and \ref{fig:phasePi2}, show the Majorana number as a function of both the chemical potential $\mu$ and magnetic impurity strength or Zeeman splitting $V_z$ for different distances between magnetic chain and the QSHI edge. We compute the Majorana number for a system with $N_x \cdot N_y=480$ sites and chains with $N_x=8$ sites for both the ferromagnetic ($\theta=0$) (Fig.~\ref{fig:phase0}) and four-site periodic spiral chains ($\theta=\pi/2$) (Fig.~\ref{fig:phasePi2}) always keeping the transverse direction fixed with $N_y\!=\!60$ sites. As a consistency check, we consider the cases where the chain is located in the central part of the sample, as shown in Figs.~\ref{fig:phase0}(d) and \ref{fig:phasePi2}(d). As expected, these systems are nearly indistinguishable from the phase diagrams resulting from a full-bulk calculation~\cite{Teixeira:PhysRevB.99.035127:2019}.

As we move the chain closer to the edge, the topological phase diagram changes, significantly increasing the  phase-space area of the topological phase for $\mu \lesssim 0.4t$ see, e.g., Figs.\ \ref{fig:phase0}(a) and \ref{fig:phasePi2}(a). 
In addition, the effect of the location of the chain relative to the edge on the phase diagram is short ranged: it all but disappears for chains located at distances larger than $3a-4a$ from the edge. 
These findings are consistent with the dependence of the QSHI phase on the chemical potential $\mu$ and the properties of the QSHI edge states. Indeed, our simulations for the Kane-Mele model show that  the system is metallic already at $\mu \sim 0.5t$, and we consequently do not see much change in the phase diagrams beyond these high doping levels. There is also no significant difference between bulk and edge chains in this metallic regime. 
In addition, the negligible effect of distance from the edge once it is larger than $4a$ is consistent with the exponential decay of the QSHI edge states wave function with the distance to the edge: for $\mu=0.4t$, the edge states extend only up to $\sim 3a$ away from the edge.
This maximum distance changes depending on the magnetic ordering along the chain, from $ \sim 1.5a$ for the ferromagnetic order ($\theta\!=\!0$, see Figs.~\ref{fig:phase0}(a-b)) to  $\sim 3a$, for the spiral order ($\theta\!=\! \pi/2$, Figs.~\ref{fig:phasePi2}(a-c)), further strengthening the notion that the QSHI edge has only a very  limited range influence.

 In addition, we also find that the shapes of the phase diagrams are at least partially determined by the properties of single impurity, as we showcase further in Appendix \ref{Sec:Single}. This is especially true for the ferromagnetic case. For a single magnetic impurity in a gapped superconductor there are always in-gap states, so-called Yu-Shiba-Rusinov-ype (YSR) states \cite{Shiba:PTP:1968,Yu:Acta:1965,Rusinov:Zh:1969,Balatsky:RevModPhys.78.373:2005}, located symmetrically around zero energy. The energy of these states depends on the magnetic impurity strength, $V_z$, and at a critical strength the energy levels cross each other at zero energy, marking a quantum phase transition. We find that the line that describes the quantum phase transition in $V_z$ vs.~$\mu$ space is in fact remarkably similar to the lower boundary of the phase diagrams in Fig.~\ref{fig:phase0}, with the same dependence on the edge location relative to the impurity. 
 
 While the results in Appendix \ref{Sec:Single} shows that YSR states are behind some features of the phase diagram for ferromagnetic chains, we note that the phase diagram is more complex in the spiral case. Still, it is possible to find an effective expression for the $(\mu, V_z)$ dependence of the YSR band crossing, which  describes the position-dependent boundary also of Fig.~\ref{fig:phasePi2}. Such a phenomenological approach is presented in Appendix~\ref{Sec:YSRBandCrossings}, and accounts for  the origin of both the phase-diagram boundary crossings and the existence of a boundless region \cite{Teixeira:PhysRevB.99.035127:2019}. While the former is mostly accidental and associated with the crossings of two pairs of YSR bands at the same point, the latter is related to the linear density of states, characteristic of a 2D Dirac systems. 

Finally, we note that setting the position-dependent superconducting order parameter, $\Delta_i$, to a constant value does not produce any significant differences in the overall phase diagrams for the ferromagnetic chain. As discussed in Appendix \ref{Sec:SOP}, setting  $\Delta_i$ as equal to the (position-independent) bulk value does not alter significantly the boundaries in the topological diagram. Moreover, we find that this is valid also for chains away from the edge. These findings, together with the persistence of the upper $V_z$ boundary in all phase diagrams, lead us to believe that the shape of the upper boundary is mainly determined by the coupling between the magnetic chain and QSHI states.

  \section{Majorana bound states}
\label{sec:MBS}

The existence of topological phases in the system with PBCs is a clear indicator of the presence of MBSs in finite chains. To explore this we perform additional calculations with OBCs and chains of different lengths. We start with relatively small chains: 20 sites for the ferromagnetic case and 21 sites for the spiral case. In both cases, the chain is embedded in a lattice with 40 sites along the zigzag (edge) direction ($N_x$) and 60 sites in the  armchair direction. We checked the consistency of these calculations by increasing the number of sites to 60 (61)  for the ferromagnetic (spiral) cases, embedding the chains in lattices with sizes $N_x=80$ and $N_y=60$. 

\subsection{Low-lying spectrum}
\label{sec:Spectrum}

\begin{figure}[t]
    \begin{center}
        \includegraphics[width=0.8\columnwidth]{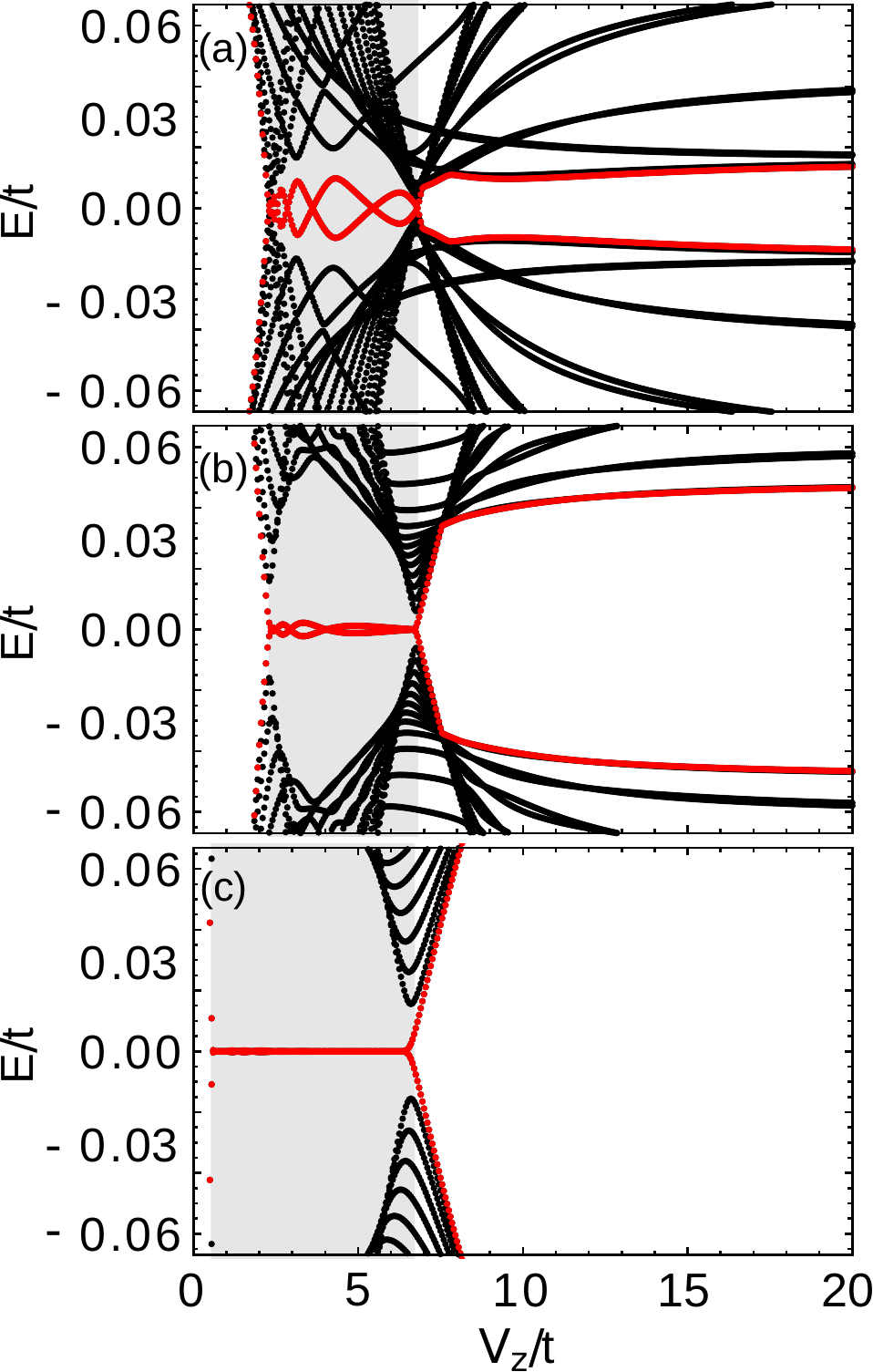}
        \caption{Ferromagnetic chain ($\theta=0$) with $\mu=0.4t$ and a fully embedded 20-site chain at (a) bulk, (b) $3a$ away from the edge and (c) edge. States closest to zero energy are marked in red and shaded regions mark the $V_z$ range of the topological phase ($\mathcal{M}=-1$) in Figs.~\ref{fig:phase0}(a), \ref{fig:phase0}(c), \ref{fig:phase0}(d). 
        }
        \label{fig:ferromagneticSpectrum}
    \end{center}
\end{figure}

We start by analyzing the energy spectrum for finite magnetic chains embedded at different distances from the edge. In Fig.~\ref{fig:ferromagneticSpectrum} we show the low-lying spectrum for $\mu=0.4t$ for ferromagnetic chains located at three different positions relative to the edge. The shaded region marks the topological phase obtained in the PBC calculations see Figs.~\ref{fig:phase0}(a), ~\ref{fig:phase0}(c) and ~\ref{fig:phase0}(d). The MBSs appear as states at or near zero energy (marked in red in the shaded region).  

At a first glance, using a  finite-length chain and OBCs (instead of PBCs) does not play an important role for a ferromagnetic chain located in the bulk. This can be seen in Fig.~\ref{fig:ferromagneticSpectrum}(a), which shows the MBSs occurring between bulk energy gap closings in the spectrum at $V_z/t \approx 2$ and $V_z/t \approx 7$, values which closely resemble those in the phase diagram of Fig.~\ref{fig:phase0}(d) calculated using PBCs. The figure also shows how the energy of the two MBSs oscillates as a function of magnetic impurity strength $V_z$. These MBS energy oscillations (MBS oscillations for short) are more prominent for shorter chains as short chains give both more overlap between the two end-point MBSs \cite{Nadj-Perge:Science:602--607:2014, Pawlak:NpjQuantumInformation:2:16035:2016,  DasSarma:Phys.Rev.B:220506:2012} and to in-gap states living along the chain \cite{Theiler19}.

As the chain is positioned closer to the edge, the main features of the bulk remain, i.e.,~the MBS oscillation pattern and the boundaries of the topological region. There are, however, important differences. First, the MBSs at  zero energy are much better characterized: the energy oscillations as a function of $V_z$ arising are heavily suppressed when the chain is positioned at the edge (Fig.~\ref{fig:ferromagneticSpectrum}(c)) as compared to the case where it is in the bulk  (Fig.~\ref{fig:ferromagneticSpectrum}(a)). Notice that the chains are of the \emph{same length} and therefore this is a very important effect of the QSHI edge. 
We also note that, in contrast to the case of a chain located in bulk of the QSHI+SC system, a finite length of the chain slightly  changes the end of the topological phase in terms of $V_z$ for edge QSHI chains: the gap opening (which marks the end of the topological phase) occurs at a smaller value ($V_z/t \approx 6.4$) relative to that calculated with PBCs shown in Fig.~\ref{fig:phase0}(c), $V_z/t \approx 6.7$.

A key difference between chains at the QSHI edge and in the QSHI bulk is the increase of the size of the superconducting gap as the chain is moved from the bulk (Fig.~\ref{fig:ferromagneticSpectrum}(a)) to the edge (Fig.~\ref{fig:ferromagneticSpectrum}(c)) of the QSHI+SC system. This change is driven by the increase of the superconducting order-parameter $\Delta_i$ when moving towards the edge of the QSHI, since there is more low-energy density of states  for a given value of the chemical potential. This increase in order parameter strength is substantial, reaching several orders of magnitude in some cases. We have already concluded that this increase does not  significantly change the onset of the topological phase (see discussion at the end of Sec.~\ref{sec:TopPhase}), but here we find that it does play an important role in localizing the MBSs in shorter chains. One of the signatures is the amplitude decrease of MBS oscillations: while such oscillations are clearly visible in Fig.~\ref{fig:ferromagneticSpectrum}(a), they are all but gone in Fig.~\ref{fig:ferromagneticSpectrum}(c) (see Appendix~\ref{Sec:SOP} for an more in-depth comparison).

\begin{figure}[t]
    \begin{center}
       \includegraphics[width=0.8\columnwidth]{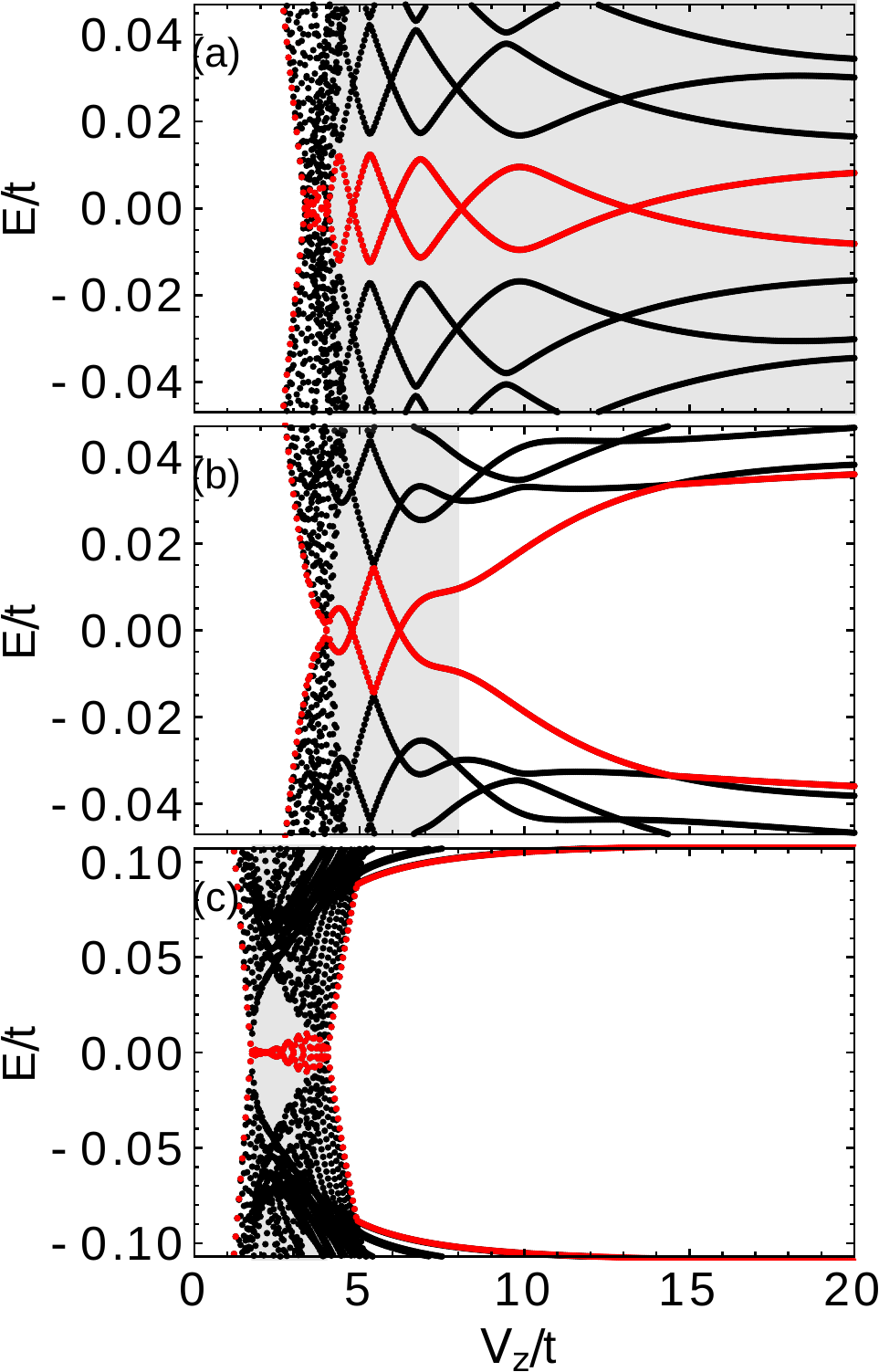}
         \caption{Spiral chain ($\theta=\pi/2$) with $\mu=0.35t$ and a fully embedded 21-site chain at (a) bulk, (b) $3a$ away from the edge and (c) edge. States closest to zero energy are in red, 
            and shaded regions mark the $V_z$ range of the topological phase ($\mathcal{M}=-1$) from Figs.~\ref{fig:phasePi2}(a), \ref{fig:phasePi2}(c), \ref{fig:phasePi2}(d).
        }
        \label{fig:ferromagneticSpectrumPi2}
    \end{center}
\end{figure}

The case with spiral magnetic order case ($\theta=\pi/2$), shown in Fig.~\ref{fig:ferromagneticSpectrumPi2}, is even more interesting. All the effects mentioned above are, in a sense, magnified. First, as shown in the phase diagrams of Fig.~\ref{fig:phasePi2}, for a given value of $\mu$ (say, $\mu=0.35t$) we can go from a ``boundless'' topological phase (no maximum value of $V_z$) to a ``upper bounded'' one by simply positioning the chain closer to the edge. 
This entails an important change of MBSs' behavior, as illustrated in Fig.~\ref{fig:ferromagneticSpectrumPi2}, which shows the low-lying spectrum for $\mu=0.35t$ and different locations of the chain (bulk, near the edge, at the edge) within the system. 
For chains located at the edge (``upper bounded'' topological regime, Fig.~\ref{fig:ferromagneticSpectrumPi2}(c)), the spectrum shows similar features as those shown in Figs.~\ref{fig:ferromagneticSpectrum}(a) and ~\ref{fig:ferromagneticSpectrum}(b) for the ferromagnetic case: an accumulation of low-energy bulk states (in black) at both the onset and end of the topological regions, marking a bulk phase transition.
By contrast, for chains located in the bulk (Fig.~\ref{fig:ferromagneticSpectrumPi2}(a)),  this value of $\mu$ corresponds to a topological phase with no upper boundary (see Fig.~\ref{fig:phasePi2}(d)) and we see no second bulk phase transition at larger $V_z$.

In between, these two extremes, a blurry picture emerges when the chain is placed near (but not exactly at) the edge. As shown in Fig.~\ref{fig:ferromagneticSpectrumPi2}(b), some properties of the spectrum are similar to those found in the bulk such as large MBS oscillation amplitudes relative to the size of the gap, indicating strong finite-size effects. On the other hand, the curvatures of the energy spectrum at the first gap closing (which marks the onset of the topological phase) are very different from the bulk and edge cases, suggesting a change in some additional effective parameter~\cite{Pan:PhysRevB.99.054507:2019}.

\subsection{Majorana polarization}
\label{sec:MP}

To gain a better understanding of effects on the MBSs by placing the magnetic chain on a QSHI edge, we can calculate the normalized \emph{Majorana polarization} (MP) defined as~\cite{Sticlet:PhysRevLett.108.096802:2012,Sedlmayr:PhysRevB.92.115115:2015}

\begin{equation}
\label{eq:majpol}
{\rm MP} = \frac{|\sum_{j\in\mathcal{R}}2 u_{j,\uparrow}v_{j,\uparrow}+2u_{j,\downarrow}v_{j,\downarrow}|}{\sum_{j\in\mathcal{R}}\langle\Psi_j|\Psi_j\rangle},
\end{equation}
where the sums over $j$ run over only sites located in \emph{half} of the system (one polarization measure per MBS),  $u_{j,\sigma} \ (v_{j,\sigma})$ is the electronic (hole) part of the negative near-zero state at site $j$ and spin $\sigma$, and $|\Psi_j\rangle=\left( u_\uparrow, u_\downarrow, v_\uparrow, v_\downarrow\right)^{\top}$ is the eigenvector of the same negative near-zero state at site $j$. Thus, a perfectly particle-hole symmetric MBS will have ${\rm MP} = 1$ and a normal state (either electronlike or holelike) has ${\rm MP} = 0$. In between these values lie the near-zero-energy bound states with ``imperfections'' in the particle-hole conjugation (the ``quasi-MBSs''). Overall, this gives a quantitative measure of the ``quality'' of the low-lying states in the topological regime: the higher the ${\rm MP}$, the closer the state is to a ``true'' MBS.

\begin{figure}[t]
	\begin{center}
		\includegraphics[width=\columnwidth]{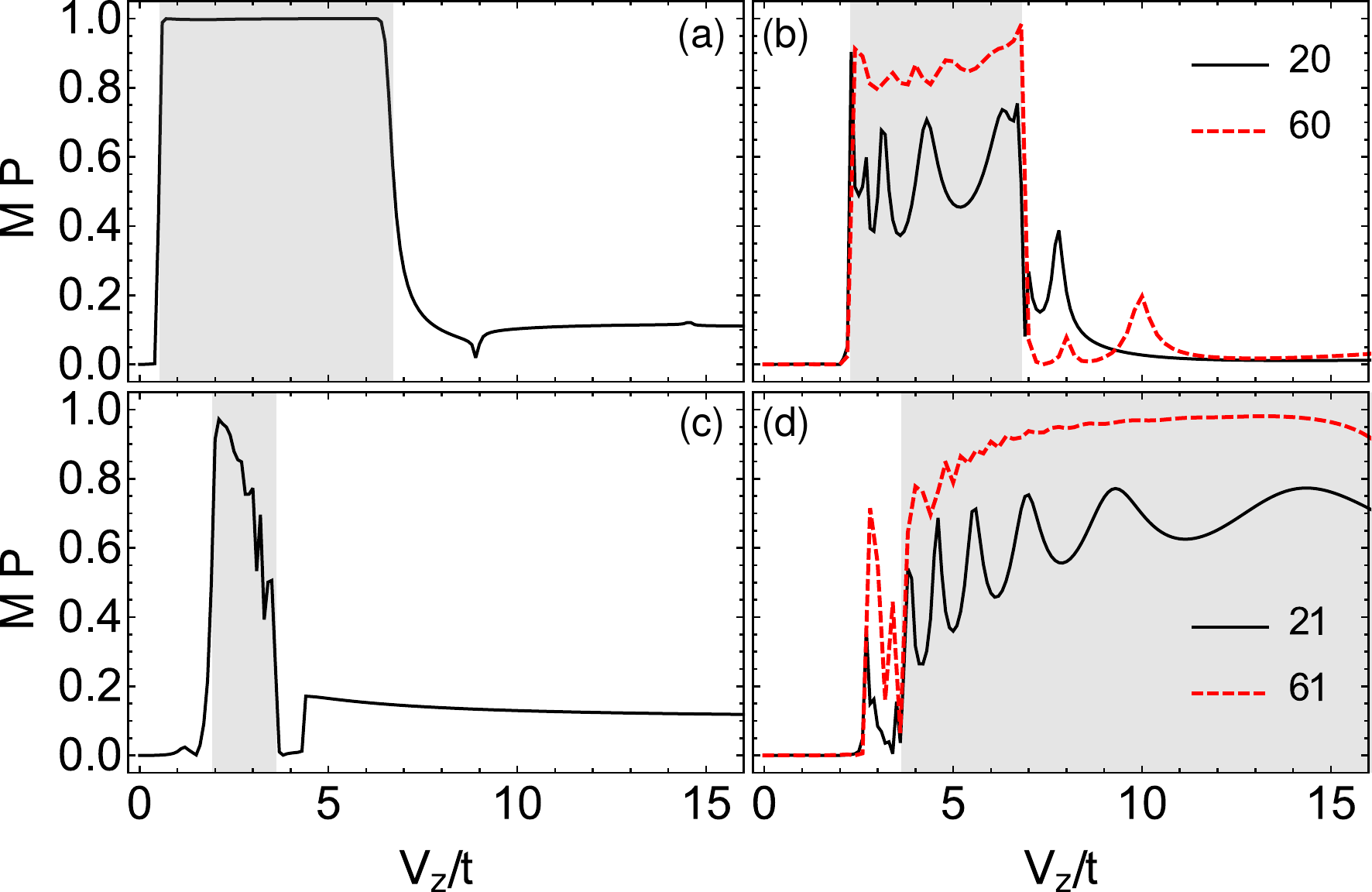}
		\caption{Majorana polarization for a ferromagnetic chain, $\theta=0$, (a), (b), and spiral, $\theta=\pi/2$, (c), (d), with  $\mu=0.4t$. The ferromagnetic chain is located at (a) edge with 20 sites and (b) bulk with 20 sites (black line) and 60 sites (red dashed). The spiral chain is located at (c) edge with 21 sites and (d) bulk with 21 sites (black line) and 61 sites (red dashed). The topological phase is inside the gray region.}
		\label{fig:MP}
	\end{center}
\end{figure} 

Figure \ref{fig:MP}(a) shows the MP versus $V_z$ for a ferromagnetic chain with 20 sites located at the QSHI edge at $\mu=0.4t$. The topological region shows an example of a ${\rm MP} = 1$ MBS. Notice the  rapidly declining from ${\rm MP}=1$ to ${\rm MP}\approx0.5$ at $V_z/t \approx 6.4$. These are the finite size effects which  cause the gap to open slightly before the topological phase transition takes place, as discussed in Sec.\ \ref{sec:MBS}. For larger values of $V_z$, the MP further declines from 0.5 to 0, which corresponds to the cone-like ABS state in the spectrum in the region $6.8t \lesssim V_z \lesssim 8.9t$ (see Fig. \ref{fig:ferromagneticSpectrum}(c)). This suggests that, while morphing from MBSs to ABSs, the system retains information from the topological phase.

  \begin{figure}[t]
  	\begin{center}
  		\includegraphics[width=\columnwidth]{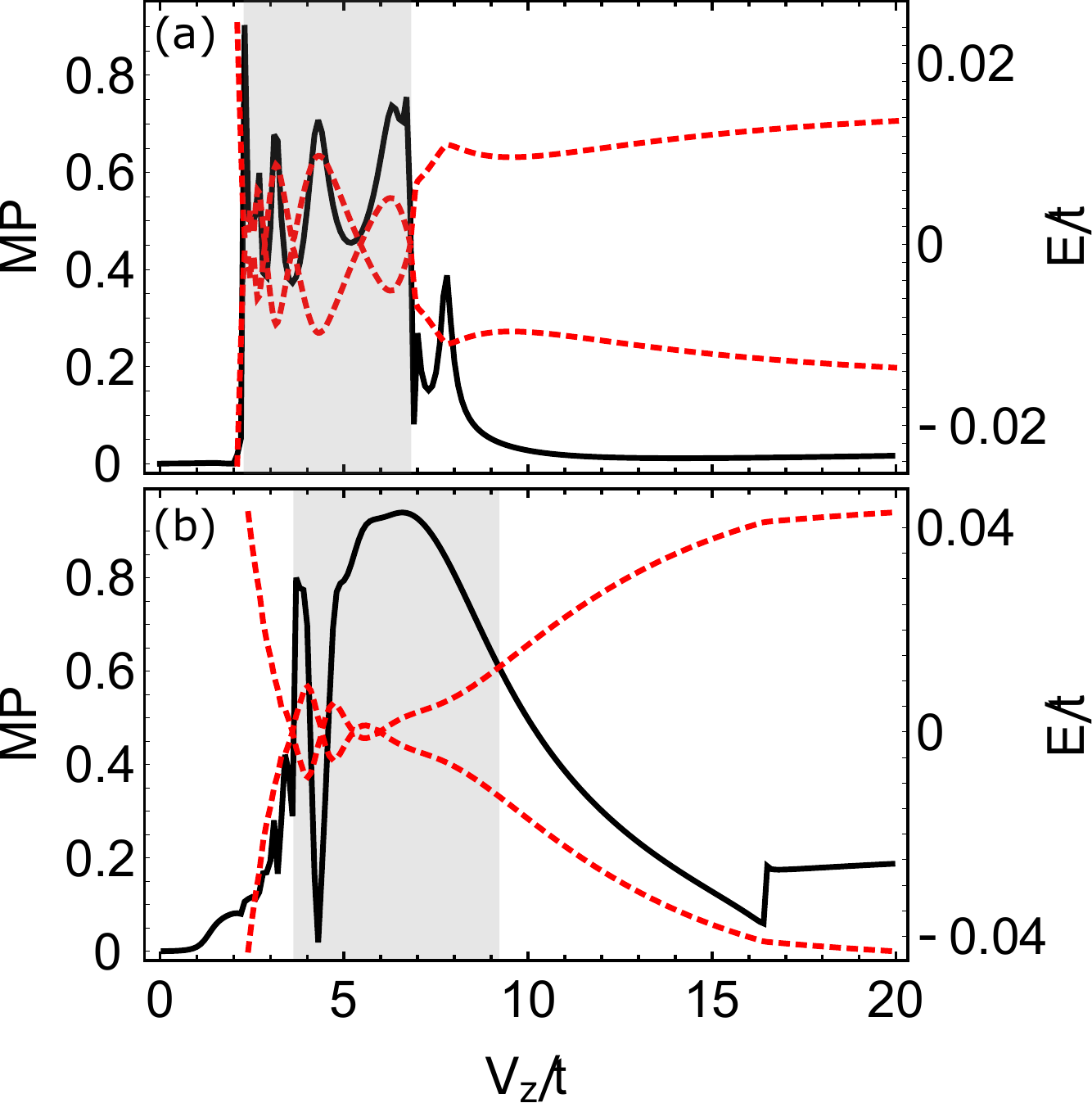}
  		\caption{Majorana polarization (thick black line) and sub-gap energy levels (dashed red line) for $\mu=0.4t$ and (a) ferromagnetic chain, $\theta=0$, with 20 sites in the bulk and (b) spiral magnetic chain, $\theta=\pi/2$,  with 21 sites at $3a$ away from the edge. Note that peaks of MP corresponds to peaks in MBS oscillation. The kink in the spectrum at (a) $V_z\approx7.9t$ and (b) $V_z\approx16.5t$, also produces a peak in MP even though it is an ABS. The topological phase lies inside the gray region.}
  		\label{fig:MPosc}
  	\end{center}
  \end{figure}

As we move the chain away from the edge to the bulk, the MP overall decreases and oscillates more, as shown in Fig.~\ref{fig:MP}(b). The maximum MP value now reaches only about 0.7, a clear decrease in ``quality'' of the MBS as compared to the case of the chain located at the edge. In this situation, increasing the chain length to 60 sites (dashed red line in Fig.~\ref{fig:MP}(b)) results in an increase of the ${\rm MP}$ maximum values to approximately 0.9, while the oscillation amplitude also reduces substantially. Only for very long chains located in the bulk do we obtain MP values close to 1, indicating the significance of finite-size effects for chains in the bulk. This is in clear contrast to the case where the chain is located at the edge, where 100\% Majorana polarization is obtained already for short chains.

We also computed the ${\rm MP}$ for the spiral order case near the boundless region, $\mu=0.4t$. The results (Fig~\ref{fig:MP}(c) and ~\ref{fig:MP}(d)) show a similar trend as in the ferromagnetic case: even relatively small ($\sim 20$ sites) chains located at the edge show ${\rm MP}$ values significatively  larger  as compared to chains in the bulk. Likewise, the transition from ABS-dominated (small ${\rm MP}$ values) to MBS-dominated (large ${\rm MP}$ values) regimes is much sharper for chains located at the edge (see Fig~\ref{fig:MP}(c)).

The  MBSs' ``low quality'' (as measured by the MP) for chains located in the bulk of the QSHI+SC system is evidenced not only by the rather low maximum values of MP (about 70\%)  but also by the rather large amplitude of the MP oscillations. Such MP oscillations are consistent with the energy versus $V_z$ oscillations seen in the MBS energy spectrum in Fig. \ref{fig:ferromagneticSpectrum}. In fact, the ${\rm MP}$ vs. $V_z$ oscillations occur \emph{in phase} with the MBS energy oscillations, as shown in Fig.~\ref{fig:MPosc}(a). Unexpectedly, the peaks in the MBS energy oscillations correspond to peaks in the MP oscillations. This is the best Majorana quality is achieved at finite energy, not at zero-energy. This in-phase oscillation is a specific behavior of the ferromagnetic chain in this system, and is not present in the spiral magnetic case (Fig.~\ref{fig:MPosc}(b)).

In addition, the MP can be used to understand how MBSs become ABSs, present for example in Fig.~\ref{fig:ferromagneticSpectrumPi2}(b) where there is a smooth transition between MBS and ABS across the bulk topological phase transition \footnote{These non-topological low-lying states are localized in the chain and evolve to MBSs at the topological transition. Thus, we refer to these as ``Andreev bound states'' even though in some cases their energies might very close to the gap edge.}. This transition does not translate immediately in the decline of MP as seen in Fig.~\ref{fig:MPosc}(b). Indeed the MP has the highest value, 0.94 at $V_z=6.7t$, after the clear deviation of the lowest-energy level from zero in Fig.~\ref{fig:ferromagneticSpectrumPi2}(b). While at the expected topological phase transition, $V_z=9.2t$, the MP has already decreased to around 0.6, we thus see a slow transition from a MBS to an ABS. This near-zero energy state finally joins the bulk spectrum at $V_z=16.5t$, where also the MP jumps to 0.2.

  \section{Concluding remarks}
  \label{sec:Conclusions}

To conclude, we investigated the properties of Majorana bound states located at the ends of a chain of magnetic impurities positioned at the edge of a 2D topological insulator with proximity-induced $s$-wave superconductivity. Our main finding is that the coupling of chain and topological insulator edge states and the enhancement of the local superconducting order parameter induced in the chain, results in a stronger exponential localization of the MBSs (and, thus, in a substantial decrease in the overlap between them) even for relatively small chain lengths. 

Placing the chain at the edge (as opposed to the bulk) of the QSHI  leads to two main effects: (i) the size of the topological region in the doping vs.~magnetic impurity strength phase diagram significantly increases for small doping levels in the QSHI regime, and (ii) there is a significant increase in the overall quality of the MBS, as measured both by the decrease in amplitude of oscillations around zero energy and by the increase in MP in the topological region.  These improvements can be understood as the result of several factors, including the changes in the spectrum of the chain states due to the coupling with QSHI edge states, as well as the increase of the superconducting order parameter near the QSHI edge.  
  
The changes in the topological phase diagram for ferromagnetic chains located either at the bulk or at the edge are driven by two distinct contributions. The lower $V_z$ boundary arises from the zero-energy crossing of in-gap magnetic states (similar to YSR states of single magnetic impurities) while the upper $V_z$ boundary is defined by zero-energy crossings of states arising from a coupling of chain states with QSHI states. In this sense, the increase in the overall topological region for chains located at the edge comes mostly from the change in the shape of the lower boundary of the phase diagram. This gives us a useful phenomenological picture, which, in principle, can also be applied to the more complex case of chains with more generic spiral magnetic ordering.

As expected, finite chains display MBSs at its ends when $\mu$ and $V_z$ are tuned to the topological regions of the phase diagram. Interestingly, the increase in the induced superconducting order parameter close to the edge of the QSHI (shown by our self-consistent calculations) leads to a stronger localization of the MBSs when a chain of a given length is moved from the bulk to the edge of the QHSI. Such ``quality improvement'' of the MBS can be quantified by the Majorana polarization (MP): for chains located at the edges, the topological transition is characterized by a sharp jump in the MP value with a maximum value close to $100 \, \%$. By contrast, for chains located in the bulk, only in the limit of very long chains the MP reaches values comparable to 100\%. In addition, for ferromagnetic chains, the MP has oscillations in phase with the spectrum around zero energy. Together with the spiral magnetic chain, that has the maximum MP at finite energies, it suggests the possibility of systems where actually finite energy values indicates the most robust MBS.

The proposed system of magnetic adatoms on QSHIs can be experimentally realized by using magnetic states (such as adatoms~\cite{Gonzalez-Herrero2016} or vacancies~\cite{Miranda2016}) on Kane-Mele-type honeycomb topological insulator~\cite{Hatsuda:AAAS:2018,Wu:Science:76:2018} placed on a superconductor substrate. Our results indicate that such an arrangement would give highly localized MBSs for a chain of magnetic sites located at a zigzag edge, making it an ideal platform for studying their properties with local probes such as STM~\cite{Nadj-Perge:Science:602--607:2014,Ruby:Phys.Rev.Lett.:197204:2015,Pawlak:NpjQuantumInformation:2:16035:2016,Jeon:Science:772:2017,Kim:ScienceAdvances::2018}.

  \begin{acknowledgments}
    
    The authors are grateful to Uppsala University and the Institute of Physics of University of S\~ao Paulo for the exchange program that allowed this project to be develop. R.L.T. and L.G.D.S. acknowledge support from FAPESP Grants No. 2016/18495-4 and 2019/11550-8, Capes, and CNPq (Graduate scholarship program, and Research Grants 308351/2017-7 and 449148/2014-9. D.K. and A.M.B.S. acknowledge support from the Swedish Research Council (Vetenskapsr\aa det, Grant No.~2018-03488), the Carl Trygger Foundation, the G\"{o}ran Gustafsson Foundation, the Swedish Foundation for Strategic Research (SSF), and the Knut and Alice Wallenberg Foundation through the Wallenberg Academy Fellows program.
    
  \end{acknowledgments}
\appendix

\section{YSR crossings in the single-impurity limit \label{Sec:Single}}

Additional insights on the shapes of the phase boundary lines for the ferromagnetic case (Fig.~\ref{fig:phase0}) can be drawn from calculations for \emph{single} magnetic impurities deposited either on the edge or in the bulk \cite{Glodzik:JPCM.32.235501} of the QSHI+SC system. In this limit, the subgap Andreev bound states are usually referred to as Yu-Shiba-Rusinov states (YSR)  and a quantum phase transition (QPT) occurs as a result of the zero-energy crossing of YSR states in the spectrum~\cite{Balatsky:RevModPhys.78.373:2005,Glodzik:JPCM.32.235501}. Since the crossing depends both on $V_z$ and $\mu$, we can construct a critical line in the $\mu$ vs. $V_z$ diagram and compare it to the phase boundaries of the chain calculation shown in Fig.~\ref{fig:phase0}.

\begin{figure}[t]
	\begin{center}
		\includegraphics[width=0.8\columnwidth]{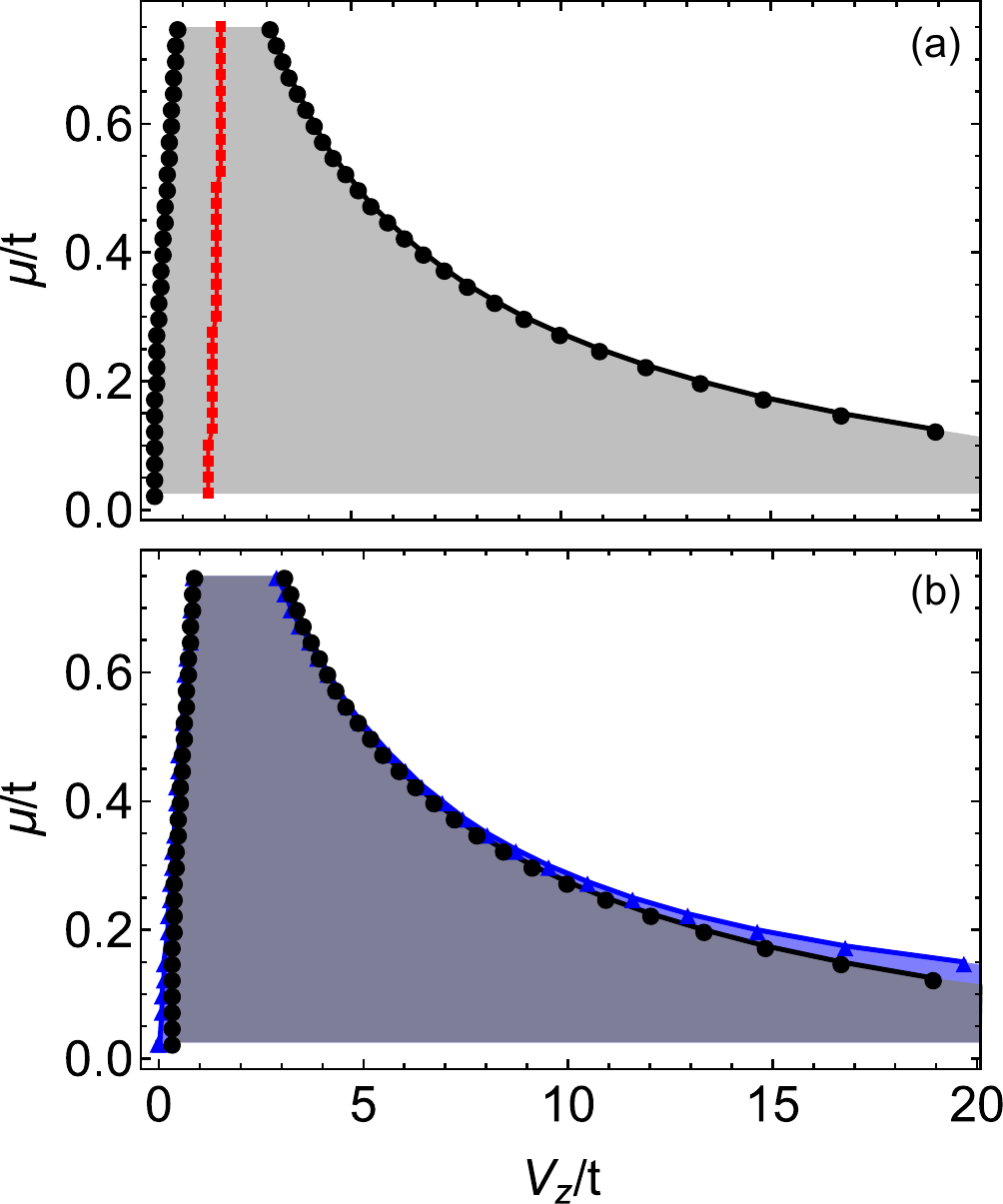}
		\caption{ Topological phase diagram for (a) single magnetic impurity located at the edge of the QSHI. The quantum phase transition line boundary is shown as red squares. The topological phase of the chain is inside the gray region (b) Ferromagnetic chain at edge of the QSHI using the bulk value $\Delta_B$ calculated at $\mu=0.025t$ (with no self-consistency). The phase diagram boundary lines are shown as blue triangles. The boundaries for the ferromagnetic phase diagram  calculated with a self-consistent $\Delta_i(\mu)$ (same as in Fig~\ref{fig:phase0}(a)) are also shown for comparison (black circles). The topological phase of the chain is inside the gray (blue) region with(out) self-consistency.}
		\label{fig:BoundaryPhases}
	\end{center}
\end{figure} 

Figure \ref{fig:BoundaryPhases} (a) shows the QPT line (in red) as a function of $V_z$ and $\mu$ for a single impurity placed at the edge of the QSHI+SC system. For completeness, we also plot the boundary lines for ferromagnetic chain phase diagram (in black) shown in Fig.~\ref{fig:phase0}.  Calculations were performed using similar parameters (namely, $N_x = 20$ sites and $N_y = 60$ sites, with the same self-consistent $\Delta_i$) as to allow a direct comparison between both systems.

The lower boundary of MBS phase diagram clearly resembles the single impurity QPT line. This underscores the close link between YSR states and MBSs in our set-up, which can provide an estimate of the lower $V_z$ boundary shape of the ferromagnetic phase diagram in similar systems involving magnetic chains on superconducting surfaces. It should be noted that a similar behavior is observed for an impurity at the bulk for which the QPT line closes resemble the lower boundary in that bulk QSHI system (not shown).

\section{Effective single-impurity model from YSR band crossings \label{Sec:YSRBandCrossings}}

The connection between the shapes of phase boundaries for the single-impurity limit  and ferromagnetic chains discussed in the previous Appendix is possible due to the uniaxial spin alignment in both cases. As such, it cannot be readily applied to chains with spiral magnetic order for which no such alignment exists.

However, we can use the similarities between the phase diagrams for different magnetic orders to establish an effective single-impurity limit for the spiral case. The first similarity is that, in both the ferromagnetic and spiral cases, the shape of one of the boundaries is essentially independent of the chain position (see Figs.~\ref{fig:phase0} and~\ref{fig:phasePi2}). The results for the  ferromagnetic case presented in the previous section suggest that the shape of the {\it other} boundary can be associated with zero-energy crossings of single-impurity YSR states. Following up on this idea, we set out to find an effective  single-impurity limit for the YSR states using the shape of this {\it other} chain position-dependent boundary as a starting point.

The topological phase transition is driven by the YSR bands (two for each impurity) which hybridize, forming the bands shown in the spectrum (Figs.~ \ref{fig:ferromagneticSpectrum} and \ref{fig:ferromagneticSpectrumPi2}). In particular, the PBC spectrum has only four YSR bands (two independent sets of bands with opposite sign), which cross at zero energy. These zero-energy crossings then also mark the onset and offset of the topological phase. In addition, these two sets have different slopes near the zero-energy crossing and can be used to distinguish the boundaries near the phase diagram's crossing \cite{Korber:PhysRevB.97.184503,Glodzik:JPCM.32.235501}.

\begin{figure}[t]
	\begin{center}
		\includegraphics[width=0.8\columnwidth]{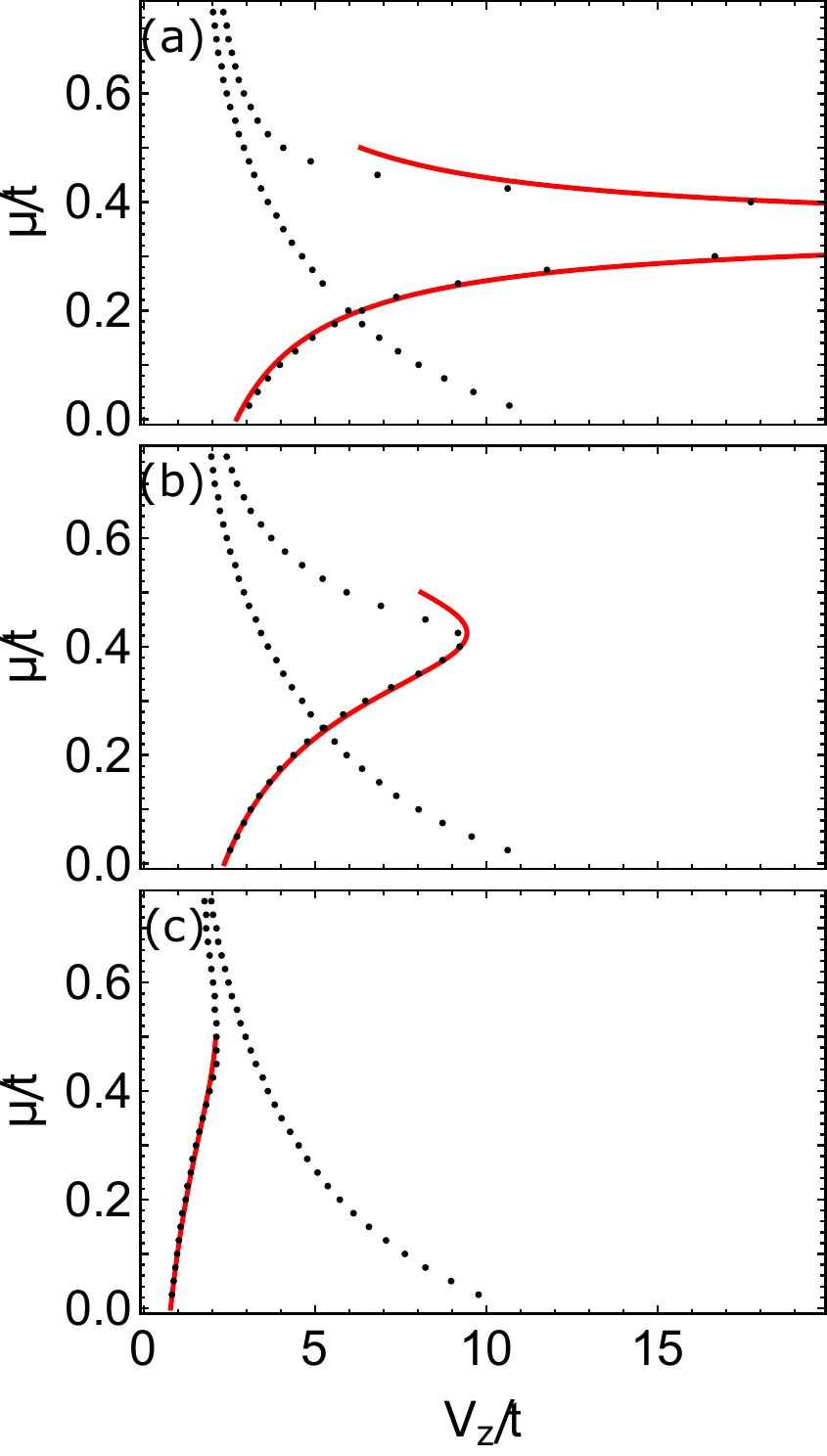}
		\caption{Phase diagram for a spiral chain ($\theta=\pi/2$), black dots, and the fit from Eqs.~(\ref{eq:VzcFit}) and (\ref{eq:VzcFitEdge}), red curve, for a chain (a) in the bulk, (b) 3a away from the edge and (c) at the edge. }
		\label{fig:FitPhases}
	\end{center}
\end{figure} 

In order to establish the effective single-impurity model, we assume the YSR bands to have energies $E_{\pm}=\pm \Delta (1-\alpha^2 V_z^2)/(1+\alpha^2 V_z^2)$ with $\alpha$ being an local impurity parameter associated with the density of states near Fermi energy~\cite{Korber:PhysRevB.97.184503}. We note that the position-dependent boundary of the bulk phase diagram, Fig~\ref{fig:phasePi2}(d), is linear with $1/V_z$. As such, we propose the following ansatz for the $\mu$-dependence of the critical  Zeeman parameter value $V_{z c}$  at the crossing:
\begin{equation}
\label{eq:VzcFit}
V_{z c}(\mu)=\sqrt{\frac{1}{\alpha^2}}= V_0+\sqrt{\frac{C_0}{(\mu-\mu_0)^2}},
\end{equation}
where $V_0$, $\mu_0$, and $C_0$ are (position-dependent) adjustable parameters. 

For chains located in the bulk, Fig.~\ref{fig:FitPhases}(a) shows that indeed the position-dependent boundary in the spiral chain phase diagram can be well fitted by Eq.~(\ref{eq:VzcFit}) up to $\mu \sim 0.45t$. The breakdown of the fitting is probably related to the QSHI to metal transition and the corresponding changes in the superconducting order parameter. 

For chains located near the edge, we can add an extra $\mu$-dependent term such that
\begin{equation}
\label{eq:VzcFitEdge}
V_{z c}(\mu)= V_0+\sqrt{\frac{C_0}{(\mu-\mu_0)^2+\beta \mu}},
\end{equation}
which takes into account the edge proximity. Again, the position-dependent boundary in the phase diagram for spiral chains can be well fitted by Eq.~(\ref{eq:VzcFitEdge}) up to $\mu \sim 0.45t$ (Figs.~\ref{fig:FitPhases}(b) and ~\ref{fig:FitPhases}(c)). 

Our phenomenological model for $V_{z c}$ is also  motivated by physical insights on the system. As in the original YSR model \cite{Balatsky:RevModPhys.78.373:2005}, the critical parameter $\alpha$ is proportional to the host density of states (DOS) at the Fermi energy $\nu(\mu)$. Even though the bulk DOS is gapped \cite{Glodzik:JPCM.32.235501}, the system behaves as  $\nu(\mu)\propto \vert \mu - \mu_{0} \vert$ near the gap edges. This can be understood as a linear approximation of the bulk DOS for energies above the QSHI gap but below the Van Hove singularity ($\lambda_{SO}\leq E\leq t$). In this region, we find $\nu(E) \propto \left( \vert E\vert-\mu^*_0 \right) \ \Theta(\vert E\vert-\lambda_{SO})$, where $\Theta$ is the Heaviside function, and $\mu^*_0=0.36t$. This value is very close to $\mu_0=0.35t$, which we find in the fit of Fig.~\ref{fig:FitPhases}(a). 
In addition, the proximity to the metallic edge states  can, as a first order approximation in the DOS, be thought of as an additional term $\beta \mu$. Finally, since we are considering YSR bands of a many-impurity system, we expect an interaction between different impurities that leads to an effective demagnetization and screening of the critical Zeeman field, such that $V^{\rm eff}_{z} = V_{z} - V_{0}$.   

The rather simple phenomenological model presented above explains several distinct features we see in the spiral phase diagram in Fig.~\ref{fig:phasePi2}. The crossings between phase boundaries occur due to accidental coincidences of two YSR crossings, each with its own dependence on the doping $\mu$. In addition, the ``upper boundless''  regime in the spiral case \cite{Teixeira:PhysRevB.99.035127:2019} can be explained by a divergence in the impurity parameter $\alpha$ at $\mu_0$, associated with the linear vanishing of the DOS discussed above.

\section{Bulk and edge superconductivity \label{Sec:SOP}}

As mentioned in Section \ref{sec:TopPhase}, the shape of the phase diagram changes noticeably as the chain is moved from the bulk to the edge of the QSHI+SC system. In particular, the ``expansion'' of the topological region in the phase diagram for small values of $\mu$ all the way down to $\mu\!=\!0$ shown in Figs.~\ref{fig:phase0}(a) and \ref{fig:phasePi2}(a) is remarkable. Since the edge states are conducting, a fair question is whether edge superconductivity is playing a role on the shape of the phase diagram or, in other words, what would be the influence of the strength of the superconducting order on the phase diagram?

\begin{figure}[b]
	\begin{center}
		\includegraphics[width=0.8\columnwidth]{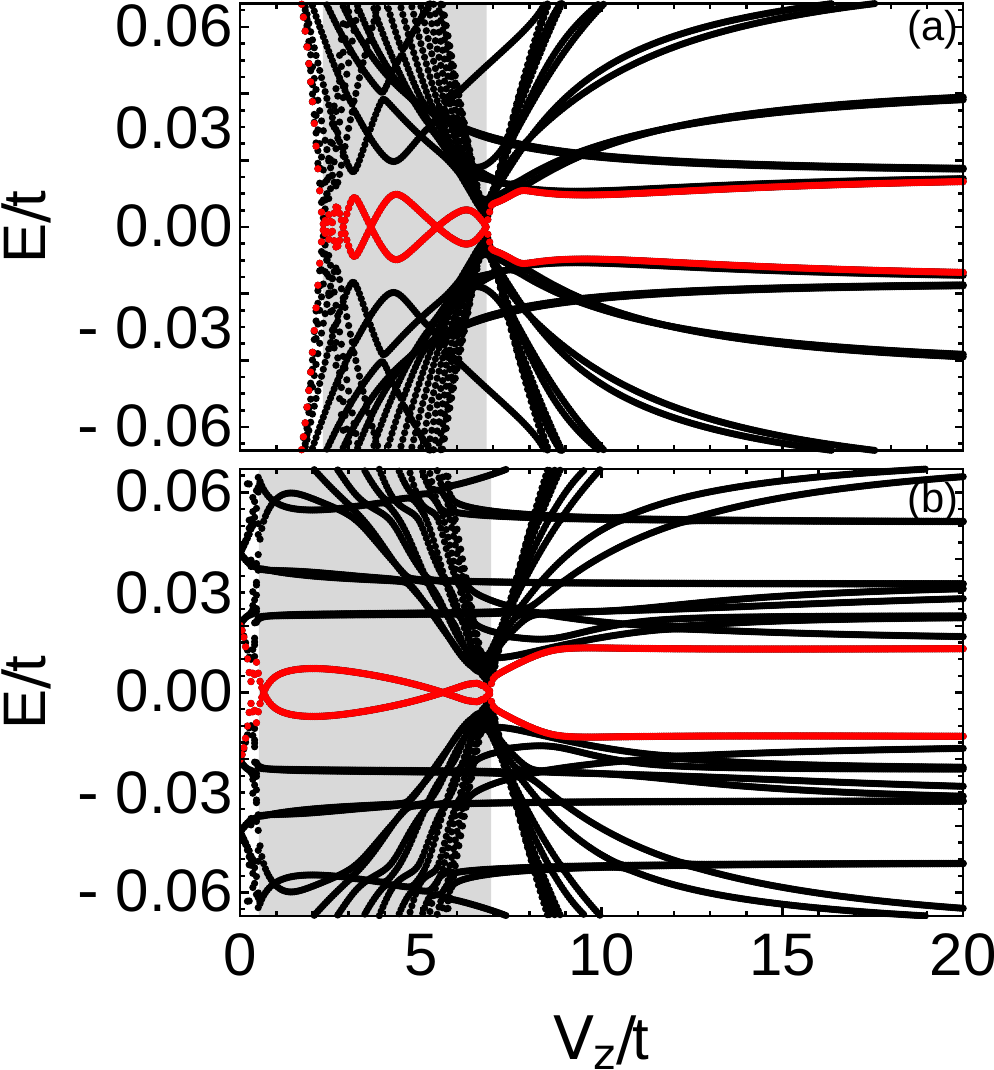}
		\caption{Ferromagnetic chain ($\theta=0$) with $\mu=0.4t$, $\Delta=\Delta_B$ and a fully embedded 20-site chain at (a) bulk and (b) edge. States closest to zero energy are in red. (a) is identical to Fig.~\ref{fig:ferromagneticSpectrum}(a)}
		\label{fig:DeltaEffect}
	\end{center}
\end{figure}

 To probe this question, we consider the extreme case where no self-consistency is used and a uniform, site-independent value of $\Delta$ taken to be equal to the \emph{bulk} value calculated for a given $\mu$.  In this scenario, we use the $\Delta$ value obtained from the self-consistent calculation for $\mu=0.025t$ for a chain  in the bulk ($\Delta_B=0.0005t$), a value much lower than the one calculated at the edge ($\Delta_E=0.22t$). Surprisingly, the overall shape of the phase diagram for the ferromagnetic chain located  at the edge of the  QSHI+SC system remains essentially unchanged from when no self-consistency is used, as shown in Fig.~\ref{fig:BoundaryPhases}(b). This indicates that edge superconductivity does not by itself drive the changes in the shapes of the phase diagram as the chain moves from the bulk to the edge of the QSHI. \footnote{A caveat is that this does not entirely apply for the spiral case as it shows extra correlations and the superconducting order parameter might play a more important role for some values of $\mu$} 

In Fig~\ref{fig:DeltaEffect}, we further explore the effect of superconducting order parameter strength by comparing the low-lying energy spectrum of the chain in the bulk and at the edge when we again artificially set the superconducting order parameter equal to the bulk value $\Delta_B$ at all sites. The amplitude of the MBS energy oscillations becomes similar between bulk and edge magnetic chains. This is in stark contrast to Fig.~\ref{fig:ferromagneticSpectrum} in the main text, which shows a difference of more than two orders of magnitude suppression of the energy oscillations for edge chains. Although $\Delta_B$ is artificially small, it is interesting to see that the main driver behind the change in energy scale for a ferromagnetic chain placed in different positions is the superconducting order parameter. However, the  change of $\Delta_i$ along the armchair direction does not explain differences in oscillation pattern, i.e., its amplitude monotonically increases/decreases or increases and then decreases, as it was observed in some cases. We expect differences in the oscillation pattern between different positions, to be associated with the changes in the spectrum QSHI edge.


%

\end{document}